\documentclass[12pt]{article}%
\usepackage{amsmath}
\usepackage{amsfonts}
\usepackage{amssymb}
\usepackage{graphicx}%
\begin{document}

\title{Generalized forms, vector fields and superspace }
\author{D C Robinson\\Mathematics Department\\King's College London\\Strand, London WC2R 2LS\\United Kingdom\\email:david.c.robinson@kcl.ac.uk}
\maketitle

\textbf{Abstract: }Vector fields with components which are generalized
zero-forms are constructed. \ Inner products with generalized forms, Lie
derivatives and Lie brackets are computed. \ The results are shown to
generalize those reported for generalized vector fields. \ Generalized affine
connections and metrics are defined and the fundamental theorem of Riemannian
geometry is extended. \ The global structure of the exterior derivative of
generalized forms is investigated.\newpage

\section{Introduction}

Generalized forms have been applied in a number of geometrically related areas
of physics. \ By extending the algebra and calculus of ordinary differential
forms new points of view about a number of different geometrical and physical
systems have been obtained. \ For example in twistor theory forms of negative
degree were introduced in order to try to extend twistor results on half-flat
space-times and to associate an abstract twistor space with general analytic
solutions of Einstein's vacuum field equations \cite{sp1} \cite{sp2}
\cite{sp3}, field theories, such as BF, Yang- Mills and gravity theories have
been reformulated as generalized topological field theories with generalized
Chern-Pontrjagin and Chern-Simons forms as Lagrangians \cite{tung1}
\cite{tung2}, \cite{rob5}, \cite{rob6} ,and generalized differential forms
have been related to forms on path space \cite{lah2}. \ Further applications
are contained in a series of papers, \cite{rob1} to \cite{rob4}, \cite{rob5}
and \cite{rob6}, devoted to the development of the formalism of generalized
forms. \ Generalized differential forms were extensively studied in these
latter papers but they dealt only with ordinary vector fields. \ Some
interesting progress going beyond ordinary vector fields was made in
\cite{lah1} and \cite{chat} where the concept of a generalized vector field
was introduced. \ In this paper the study of vector fields is continued and
their work is extended. \ First a dictionary between the algebra and calculus
of certain functions and vector fields on a superspace and the algebra and
calculus of generalized forms and vector fields is established. \ This
dictionary is not only useful in its own right but it also facilitates the
introduction of the concept of a generalized form-valued vector field. \ In
the case considered here such an object is determined by an ordered pair
consisting of of an ordinary vector field and a $\left(  1,1\right)  $ type
tensor field. \ This concept includes generalized vector fields as a special
case and provides an improved understanding of their properties. \ The
introduction of generalized form-valued vector fields also enables generalized
metrics and affine connections to be defined by constructions which can be
extended to more general connections.

A brief review of the algebra and differential calculus of generalized forms
needed in this paper is given in the second section. \ Different types of
generalized differential forms, on an $n$ dimensional manifold $M$, are
labelled by a non-negative integer $N$. \ In this paper only the case where
$N=1$ is considered but the results are easily extendible to $N\geqq2$. \ \ A
type $N=1$ generalized $p-$form is defined by an ordered pair consisting of
two ordinary forms of degrees $p$ and $p+1$ respectively, where $-1\leqq
p\leqq$ $n$. \ The module of type $N=1$ generalized $p-$forms on $M$ is
denoted $\Lambda_{(1)}^{p}(M)$. \ The exterior product for generalized forms
makes the vector space of type $N=1$ forms at a \ point $x$ in $M$,
$\Lambda_{(1)}^{\bullet}(x)$ $=\oplus_{p=-1}^{p=n}\Lambda_{(1)}^{p}(x)$, into
an associative algebra, in fact a super-commutative graded algebra.
\ Generalized forms of degree zero form a commutative ring with $1\neq0.$
\ The graded module, and super-commutative graded algebra over the ring of
smooth functions, on $M$ is equipped with an exterior derivatives,
$d:\Lambda_{(1)}^{p}(M)\rightarrow\Lambda_{(1)}^{p+1}(M)$, a super-derivation
of degree one. \ While both the exterior algebra and differential calculus
satisfied by generalized forms are similar to the algebra and calculus of
ordinary forms there are some differences. \ For instance, generalized forms
of degree $p=-1$ are allowed and the generalized de Rham cohomology can be
different from the de Rham cohomology of ordinary forms.

The actions of ordinary vector fields on generalized forms on $M$ presented
previously, \cite{rob2}, are summarized in section three. \ In the fourth
section the algebra and calculus of generalized forms and the actions of
ordinary vector fields on $M$ are represented on the Whitney sum of a reverse
parity trivial line bundle and the reverse parity tangent bundle over $M$.
\ This extends to generalized forms a known approach to ordinary differential
forms, \cite{voronov}. \ This point of view is employed in the fifth section
where generalized form-valued vector fields are introduced and their
properties\ explored. \ The definitions of the interior products and Lie
derivatives of generalized forms with respect to such vector fields and the
definition of a Lie bracket are given, extending the results of section three
from ordinary vector fields to generalized form-valued vector fields.
\ Generalized vector fields, which were introduced in \cite{lah1} and
\cite{chat} and applied to the Hamiltonian formalism for a free relativistic
particle, are discussed and shown to form a sub-class of generalized-form
valued vector fields. \ Two examples of the use of generalized form-valued
vector fields are presented, one introducing generalized form-valued
Hamiltonian vector fields. \ An application is given in the sixth section
where generalized form-valued vector fields are used in the construction of
the tensor calculus of generalized affine connections and metrics. \ The
compatibility conditions of generalized affine connections and generalized
metrics are presented and \ an extension of the fundamental theorem of
Riemannian geometry is obtained. \ The seventh section contains a brief
summary of the results and an outline of ways in which they can be used and
developed. \ Finally there is an appendix in which the global structure of
exterior derivatives of type $N=1$ forms is discussed.

The results in this paper can apply to manifolds and geometrical objects that
are real or complex but in this paper it will be assumed that the geometry is
real, all geometrical objects are smooth and $M$ is an $n-$dimensional real,
smooth, orientable and oriented manifold. \ Bold-face Roman letters are used
to denote generalized forms and generalized vector fields,\ ordinary forms on
$M$ are usually denoted by Greek letters and ordinary vector fields on $M$ by
lower case Roman letters. \ Occasionally the degree of a form is indicated
above it. \ The exterior product of any two forms, for example $\alpha$ and
$\beta,$ is written $\alpha\beta$, and as usual, any ordinary $p-$form
$\overset{p}{\alpha}$, with $p$ either negative or greater than $n$, is zero.
\ The Einstein summation convention is used.

\section{Algebra and calculus of generalized forms}

The algebraic and differential properties of generalized forms are outlined in
this section using the notation of \cite{rob5} and \cite{rob6}. \ In this
paper generalized forms will \ be expressed in terms of a minus one-form which
is linearly independent of ordinary forms on $M$, \cite{rob5}. \ Hence a basis
for type $N=1$ generalized forms consists of any basis for ordinary forms on
$M$ augmented by a minus one-form $\mathbf{m}$. \ Apart from having a degree
of minus one the latter has the same algebraic properties as an ordinary
exterior form. \ It satisfies the ordinary distributive and associative laws
of exterior algebra and the exterior product rule%
\begin{equation}
\overset{p}{\alpha}\mathbf{m}=(-1)^{p}\mathbf{m}\overset{p}{\alpha};\text{
}\mathbf{m}^{2}=0,\text{ }%
\end{equation}
together with the condition of linear independence. \ Thus, for a given choice
of $\mathbf{m}$, a generalized p-form, $\overset{p}{\mathbf{a}}\in
\Lambda_{(1)}^{p}$, can be written as%
\begin{equation}
\overset{p}{\mathbf{a}}=\overset{p}{\alpha}+\overset{p+1}{\alpha}\mathbf{m},
\end{equation}
where $\overset{p}{\alpha},$ and $\overset{p+j}{\alpha}$ are, respectively,
ordinary $p-$ and $(p+1)-$forms and $p$ can take integer values from $-1$ to
$n$. \ At a point $x$ in $M$ the generalized $p-$forms of type N=$1$,
$\Lambda_{(1)}^{p}(x)$, form a real vector space of dimension $\frac
{(1+n)!}{(1+p)!(n-p)!}$. \ The dimension of $\Lambda_{(1)}^{\bullet}%
(x)=\oplus_{p=-1}^{p=n}\Lambda_{(1)}^{p}(x)$ is $2^{1+n}$.

If $\varphi$ is a smooth map between manifolds $P$ and $M,$ $\varphi
:P\rightarrow M,$ then the induced map of type $N=1$ generalized forms,
$\varphi_{(1)}^{\ast}:\Lambda_{(1)}^{p}(M)\rightarrow\Lambda_{(1)}^{p}(P)$, is
the linear map defined by using the standard pull-back map, $\varphi^{\ast}$,
for ordinary forms%
\begin{equation}
\varphi_{(1)}^{\ast}(\overset{p}{\mathbf{a}})=\varphi^{\ast}(\overset
{p}{\alpha})+\varphi^{\ast}(\overset{p+1}{\alpha})\mathbf{m},
\end{equation}
and $\varphi_{(1)}^{\ast}(\overset{p}{\mathbf{a}}\overset{q}{\mathbf{b}%
})=\varphi_{(1)}^{\ast}(\overset{p}{\mathbf{a}})\varphi_{(1)}^{\ast}%
(\overset{q}{\mathbf{b}})$. \ Hence $\varphi_{(1)}^{\ast}(\mathbf{m)=m}$.

Henceforth in this paper, in addition to assuming that the exterior derivative
of generalized forms satisfies the usual properties, it is\ assumed that%
\begin{equation}
d\mathbf{m}=\epsilon,
\end{equation}
where $\epsilon$ denotes a real constant. \ If $\mathbf{m\mapsto}%
\widetilde{\mathbf{m}}\mathbf{=\mu m}$, where $\mu$ is a non-zero function on
$M$, then $\overset{p}{\mathbf{a}}=\overset{p}{\alpha}+\overset{p+1}{\alpha
}\mathbf{m}=\overset{p}{\alpha}+\overset{p+1}{\widetilde{\alpha}}%
\widetilde{\mathbf{m}}$, where$\overset{p+1}{\widetilde{\alpha}}=\mu
^{-1}\overset{p+1}{\alpha}$.\ \ \ Furthermore d$\widetilde{\mathbf{m}%
}\mathbf{=}\widetilde{\epsilon}$, where $\widetilde{\epsilon}$ is also a real
constant, if and only if $d\mu=0$ and then $\widetilde{\epsilon}=\mu\epsilon$.

The exterior derivative of a type $N=1$ generalized form $\overset
{p}{\mathbf{a}}$ is then%
\begin{equation}
d\overset{p}{\mathbf{a}}=[d\overset{p}{\alpha}+(-1)^{p+1}\epsilon\overset
{p+1}{\alpha}]+d\overset{p+1}{\alpha}\mathbf{m},
\end{equation}
where $d$ is the ordinary exterior derivative when acting on ordinary forms.
\ The exterior derivative $d:$ $\Lambda_{(1)}^{p}(M)\rightarrow\Lambda
_{(1)}^{p+1}(M)$ is an anti-derivation of degree one,%
\begin{align}
d(\overset{p}{\mathbf{a}}\overset{q}{\mathbf{b}})  &  =d\overset{p}%
{\mathbf{a}}\overset{q}{\mathbf{b}}+(-1)^{p}\overset{p}{\mathbf{a}}%
d\overset{q}{\mathbf{b}},\\
d^{2}  &  =0.\nonumber
\end{align}
and $(\Lambda_{(N)}^{\bullet}(M),d)$ is a differential graded algebra. \ The
exterior derivative is discussed in more detail in the appendix.

\section{Vector fields and type $N=1$ forms}

In this section the definitions of the inner product and Lie derivative of
type $N=1$ forms by ordinary vector fields introduced in \cite{rob2} will be
summarized. \ Let $v$ be an ordinary vector field tangent to M, $v\in
\mathcal{V}_{(o)}(M)$, where $\mathcal{V}_{(o)}(M)$ is the module of ordinary
vector fields over $C^{\infty}(M)$, the real valued functions on $M$. \ \ Let
the generalized p-form $\overset{p}{\mathbf{a}}$ and q-form $\overset
{q}{\mathbf{b}}$ be given, respectively, by $\overset{p}{\alpha}+\overset
{p+1}{\alpha}\mathbf{m}$ and $\overset{q}{\beta}+\overset{q+1}{\beta
}\mathbf{m}$. \ The inner product or contraction operator on generalized
forms, $i_{v}:\Lambda_{(1)}^{p}\rightarrow\Lambda_{(1)}^{p-1}$, for $-1\leqq
p\leqq n$, is defined in terms of the inner product for ordinary forms by
\begin{equation}
i_{v}\overset{p}{\mathbf{a}}=i_{v}\overset{p}{\alpha}+(i_{v}\overset
{p+1}{\alpha})\mathbf{m}.
\end{equation}
Since $\overset{-1}{\alpha}=0,$ and $i_{v}\overset{0}{\alpha}=0$,
$i_{v}\overset{-1}{\mathbf{a}}=0$ and $i_{v}\overset{0}{\mathbf{a}}%
=(i_{v}\overset{1}{\alpha})\mathbf{m}$. \ Furthermore, for any two vector
fields $v$ and $w\in\mathcal{V}_{(o)}(M)$%
\begin{equation}
i_{w}(i_{v}\overset{p}{\mathbf{a}})+i_{v}(i_{w}\overset{p}{\mathbf{a}})=0.
\end{equation}
It is a straight forward matter to show that Eq.(7) implies that%
\begin{equation}
i_{v}(\overset{p}{\mathbf{a}}\overset{q}{\mathbf{b}})=(i_{v}\overset
{p}{\mathbf{a}})\overset{q}{\mathbf{b}}+(-1)^{p}\overset{p}{\mathbf{a}}%
(i_{v}\overset{q}{\mathbf{b}}),
\end{equation}
that is%

\[
i_{v}(\overset{p}{\mathbf{a}}\overset{q}{\mathbf{b}})=i_{v}(\overset{p}%
{\alpha}\overset{q}{\beta})+[i_{v}(\overset{p}{\alpha}\overset{q+1}{\beta
})+(-1)^{q}i_{v}(\overset{p+1}{\alpha}\overset{q}{\beta})]\mathbf{m},
\]
The Lie derivative with respect to $v$, $\pounds _{v}$ is defined by
\begin{equation}
\pounds _{v}\overset{p}{\mathbf{a}}=i_{v}d\overset{p}{\mathbf{a}}%
+d(i_{v}\overset{p}{\mathbf{a}}),
\end{equation}
from which it follows that $d\circ\pounds _{v}\overset{p}{\mathbf{a}%
}=\pounds _{v}\circ d\overset{p}{\mathbf{a}}$.

A calculation then shows that
\begin{equation}
\pounds _{v}\overset{p}{\mathbf{a}}=\pounds _{v}\overset{p}{\alpha
}+(\pounds _{v}\overset{p+1}{\alpha})\mathbf{m}.
\end{equation}
A direct consequence of Eqs.(9) and (10) above is that $\pounds _{v}$
satisfies the Leibniz rule
\begin{equation}
\pounds _{v}(\overset{p}{\mathbf{a}}\overset{q}{\mathbf{b)}}=(\pounds _{v}%
\overset{p}{\mathbf{a}})\overset{q}{\mathbf{b}}+\overset{p}{\mathbf{a}%
}\pounds _{v}(\overset{q}{\mathbf{b}}),
\end{equation}
A couple of important differences from results for ordinary forms should be
noted,%
\begin{equation}
\pounds _{v}\overset{-1}{\mathbf{a}}=i_{v}(d\overset{0}{\alpha})\mathbf{m,}%
\end{equation}
and%
\begin{equation}
i_{v}(d\mathbf{\overset{0}{\mathbf{a}})}\mathbf{=}i_{v}d\overset{0}{\alpha
}-\epsilon i_{v}\overset{1}{\alpha}+i_{v}d\overset{1}{\alpha}\mathbf{m,}%
\end{equation}
In contrast to the case for ordinary zero-forms, the latter is not equal to
the Lie derivative for%
\begin{equation}
\pounds _{v}\overset{0}{\mathbf{a}}=i_{v}d\overset{0}{\alpha}+[i_{v}%
d\overset{1}{\alpha}+d(i_{v}\overset{1}{\alpha})]\mathbf{m.}%
\end{equation}
If $v$ and $w$ are vector fields in $M$ it follows from the definitions above
that%
\begin{equation}
(\pounds _{v}\circ i_{w}-i_{w}\circ\pounds _{v})\overset{p}{\mathbf{a}%
}=i_{[v,w]}\overset{p}{\mathbf{a}},
\end{equation}%
\begin{equation}
(\pounds _{v}\circ\pounds _{w}-\pounds _{w}\circ\pounds _{v})\overset
{p}{\mathbf{a}}=\pounds _{[v,w]}\overset{p}{\mathbf{a}},
\end{equation}
where $[v,w]$ denotes the Lie bracket of the vector fields $v$ and $w$. \ The
latter type of equation will be used in section five to define a
generalization of the Lie bracket.

In summary, with these definitions, $\Lambda_{(1)}$ is a graded algebra and
there is a natural grading of these linear operators on $\Lambda(M)$, $d$ is
degree $1$, $\pounds _{v}$ is of degree $0$ and $\iota_{v}$ is of degree $-1$.
\ These derivations span a super Lie algebra and satisfy the H.Cartan
formulae, \cite{chern},
\begin{align}
d\circ d  &  =0\text{, }i_{v}\circ i_{w}+i_{w}\circ i_{v}=0,\\
\pounds _{v}  &  \equiv d\circ i_{v}+i_{v}\circ d\Rightarrow\text{ }%
d\circ\pounds _{v}-\pounds _{v}\circ d=0.\nonumber\\
\pounds _{v}\circ\pounds _{w}-\pounds _{w}\circ\pounds _{v}  &
=\pounds _{[v,w]}\text{, }\pounds _{v}\circ i_{w}-i_{w}\circ\pounds _{v}%
=i_{[v,w]},\nonumber
\end{align}
for all vector fields $v$ and $w$ $\in\mathcal{V}_{(o)}(M)$.

\section{Representation of the algebra and calculus of generalized forms}

The algebra and calculus of ordinary differential forms on $M$ can be
expressed in terms of functions and vector fields on the reverse parity
tangent bundle, $\Pi TM$, of $M$, \cite{voronov}. \ A recent exposition
containing further references can be found in \cite{witten3}. \ A sample of
texts where superspace calculations\ are discussed is \cite{dewitt},
\cite{rogers}, \cite{wess}.

The reverse parity tangent bundle is just the ordinary tangent bundle with the
parity reversed in the fibre directions. \ If $x^{\alpha}$ , $\alpha
=1....n=\dim M$, denote local coordinates on $M$ then local coordinates on
$\Pi TM$ are obtained by adding to these $n$ anticommuting fibre coordinates.
\ The latter are obtained by replacing the natural tangent bundle fibre
coordinates with $n$ anti-commuting (fermionic) coordinates with the same
transformation properties. \ These can be denoted by the symbols $dx^{\alpha}$
or, as will be done here for notational clarity by $\zeta^{a}$. \ Then an
ordinary $p-$form $\rho$ on $M$ with coordinate basis components $\rho
_{\alpha_{1}....\alpha_{p}}(x^{\alpha})$.
\begin{equation}
\rho=\frac{1}{p!}\rho_{\alpha_{1}....\alpha_{p}}(x^{\alpha})dx^{\alpha
_{1.....}}dx^{\alpha_{p}},
\end{equation}
corresponds to, $r$, a homogeneous polynomial of degree $p$ in the
anticommuting \ fibre coordinates on $\Pi TM$%
\begin{equation}
r=\frac{1}{p!}\rho_{\alpha_{1}....\alpha_{p}}(x^{\alpha})\zeta^{\alpha
_{1.....}}\zeta^{\alpha_{p}}.
\end{equation}

The exterior product of ordinary forms $p-$ and $q-$forms on $M$ corresponds
to the product of such functions (which are homogeneous polynomials of
respective degrees $p$ and $q$ in the anticommuting coordinates) on $\Pi TM$.
\ The exterior derivative of ordinary $p-$forms on $M$, where $p>0$,
corresponds to the action of the odd vector field $\zeta^{\alpha}%
\frac{\partial}{\partial x^{\alpha}}$ on the corresponding functions on $\Pi
TM$ and the correspondence can be written as%
\begin{equation}
d\rho\leftrightarrow\zeta^{\alpha}\frac{\partial r}{\partial x^{\alpha}}.
\end{equation}
The interior product, $i_{v}$, of an ordinary $p-$ form $\rho$ on $M$ by a
vector field $v$= $v^{\alpha}\frac{\partial}{\partial x^{\alpha}}$
$\in\mathcal{V}_{(o)}(M)$, corresponds to the action of the odd vector field
$v^{\alpha}\frac{\partial}{\partial\zeta^{\alpha}}$\ on the function $r$ in
$\Pi TM$; the correspondence can be written as%
\begin{equation}
i_{v}\rho\leftrightarrow v^{\alpha}\frac{\partial r}{\partial\zeta^{\alpha}}%
\end{equation}
For example, the local coordinate expression for the action of a vector field
$v$ on a zero-form $\rho$ on $M$ is%
\begin{equation}
v(\rho)=i_{v}d\rho\text{,}%
\end{equation}
and using the correspondences above
\begin{equation}
v(\rho)\leftrightarrow v^{\alpha}\frac{\partial}{\partial\zeta^{\alpha}}%
(\zeta^{b}\frac{\partial r}{\partial x^{b}}).
\end{equation}

The Lie derivative $\pounds _{v}$ on $M$ corresponds to the even vector field,
$[d,i_{v}]$, on $\Pi TM$ where $[.,.]$ denotes the super Lie bracket of the
odd vector fields $d$ and $i_{v}$ or equivalently the supercommutator of the
differential operators on $\Pi TM$,
\begin{align}
\pounds _{v}  &  \leftrightarrow\lbrack d,i_{v}]=d\circ(i_{v})+i_{v}\circ d,\\
\lbrack d,i_{v}]r  &  =v^{\alpha}\frac{\partial r}{\partial x^{\alpha}}%
+\frac{\partial v^{\alpha}}{\partial x^{\beta}}\zeta^{\beta}\frac{\partial
r}{\partial\zeta^{\alpha}}.
\end{align}

Henceforth the same notation will be used for corresponding operators on $M$
and reverse parity bundles. \ It will be clear from the context which is
meant, for example on $\Pi TM$%
\begin{align*}
\pounds _{v} &  =[d,i_{v}]\\
\pounds _{v}r &  =v^{\alpha}\frac{\partial r}{\partial x^{\alpha}}%
+\frac{\partial v^{\alpha}}{\partial x^{\beta}}\zeta^{\beta}\frac{\partial
r}{\partial\zeta^{\alpha}}%
\end{align*}
and correspondingly on $M$%
\[
\pounds _{v}\rho=[d\circ(i_{v})+i_{v}\circ d]\rho.
\]

If $w=w^{\alpha}\frac{\partial}{\partial x^{\alpha}}$ is a another vector
field $\in\mathcal{V}_{(o)}(M)$, \ calculation of the super commutator on $\Pi
TM$ of $\pounds _{v}$ and $\pounds _{w}$ gives%
\begin{equation}
\lbrack\text{ }\pounds _{v},\pounds _{w}]=\text{ }\pounds _{v}\circ
\pounds _{w}-\pounds _{w}\circ\pounds _{v}=\pounds _{[v,w]}.
\end{equation}
Similarly computing the supercommutator on $\Pi TM$ gives%
\begin{equation}
\lbrack\pounds _{v},i_{w}]=\pounds _{v}\circ i_{w}-i_{w}\circ\pounds _{v}%
=i_{[v,w]},
\end{equation}
where $[v,w]$ is the Lie bracket of $v$ and $w$. \ By the correspondences the
same results hold for forms on $M$.

It is a straightforward matter to extend these ideas to generalized forms of
all types on $M$. \ Here only type $N=1$ forms will be considered in detail.
\ Let $\widetilde{M}$ be the Whitney sum of $\Pi TM$ and a trivial reverse
parity line bundle over $M$, that is\ a trivial line bundle with fibre
$\mathbb{R}^{1}$ replaced by $\mathbb{R}^{0\mid1}$, and let $\mathbb{R}%
^{0\mid1}$ have anti-commuting coordinate $\mu$ . \ \ Local coordinates for
$\widetilde{M}$ can then be chosen to be the commuting coordinates $x^{\alpha
}$ together with the anti-commuting coordinates $\ \zeta^{a}=dx^{\alpha}$
and\ $\mu$. \ If
\begin{equation}
\mathbf{r}=\rho+\sigma\mathbf{m}%
\end{equation}
is a generalized \ $p-$form on $M$ and the ordinary $p-$ and $(p+1)-$forms
$\rho$ and $\sigma$ have respective coordinate basis components $\rho
_{\alpha_{1}....\alpha_{p}}$ and $\sigma_{\alpha_{1}....\alpha_{p+1}}$ and%
\begin{equation}
d\mathbf{m=\epsilon,}\text{ }%
\end{equation}
where $\epsilon$ is a constant, then $\mathbf{r}$ corresponds to the function%
\begin{equation}
\mathfrak{r}=\frac{1}{p!}\rho_{\alpha_{1}....\alpha_{p}}(x^{\alpha}%
)\zeta^{\alpha_{1.....}}\zeta^{\alpha_{p}}+\frac{1}{(p+1)!}\sigma_{\alpha
_{1}....\alpha_{p+1}}(x^{\alpha})\zeta^{\alpha_{1.....}}\zeta^{\alpha_{P+1}%
}\mu
\end{equation}
on $\widetilde{M}$ . \ The exterior product of generalized forms in $M$
corresponds to the product of such functions in $\widetilde{M}$. \ \ The
exterior derivative of a generalized form $\mathbf{r}$ in $M$, $d\mathbf{r}$,
corresponds to the action of the odd vector field on the corresponding
function in $\widetilde{M}$%
\begin{equation}
d:\mathfrak{r}\rightarrow(\zeta^{\alpha}\frac{\partial}{\partial x^{\alpha}%
}+\epsilon\frac{\partial}{\partial\mu})\mathfrak{r.}%
\end{equation}
The interior product, $i_{v}\mathbf{r}$, of a \ generalized $p-$form
$\mathbf{r}$ on $M$ by a vector field $v$= $v^{\alpha}\frac{\partial}{\partial
x^{\alpha}}$ $\in\mathcal{V}_{(o)}(M)$ is the generalized $(p-1)-$form%
\begin{equation}
i_{v}\mathbf{r}=i_{v}\rho+(i_{v}\sigma)\mathbf{m}%
\end{equation}
which corresponds to $v^{\alpha}\frac{\partial\mathfrak{r}}{\partial
\zeta^{\alpha}}$ on $\widetilde{M}$, i.e. on $\widetilde{M}$
\begin{equation}
i_{v}:\mathfrak{r}\rightarrow v^{\alpha}\frac{\partial\mathfrak{r}}%
{\partial\zeta^{\alpha}}\mathbf{.}%
\end{equation}
($d\mathfrak{r}=(\zeta^{\alpha}\frac{\partial}{\partial x^{\alpha}}%
+\epsilon\frac{\partial}{\partial\mu})\mathfrak{r}$ and $i_{v}\mathfrak{r}%
=v^{\alpha}\frac{\partial\mathfrak{r}}{\partial\zeta^{\alpha}}$ in accordance
with the convention established above.)

\section{Generalized form-valued vector fields}

These ideas of the previous sections can be extended to include vector fields
with generalized form-valued components. \ \ Define such a\ type $N$ vector
field on $M$ by $\mathbf{V=v}^{\rho}\frac{\partial}{\partial x^{\rho}}$ where
the components $\mathbf{v}^{\rho}$ are type $N$ zero-forms which transform as
the components of a vector field . \ In the case considered in this paper
$N=1$ and%
\begin{equation}
\mathbf{V=v}^{\rho}\frac{\partial}{\partial x^{\rho}}=(v^{\rho}+v_{\sigma
}^{\rho}dx^{\sigma}\mathbf{m)}\frac{\partial}{\partial x^{\rho}}=v+(v_{\sigma
}^{\rho}dx^{\sigma}\mathbf{m)}\frac{\partial}{\partial x^{\rho}},
\end{equation}
where $v=v^{\rho}\frac{\partial}{\partial x^{\rho}}$. \ Hence $\mathbf{V}$ is
determined by an ordinary vector field $v$ and a $(1,1)$ type tensor field
$v_{\sigma}^{\rho}\frac{\partial}{\partial x^{\rho}}\otimes dx^{\sigma}$ on
$M$. \ The set of all such vector field in $M$, $\mathbf{V}$, is naturally a
module, $\mathcal{V}_{(1)}(M)$, over the generalized zero forms on $M$,
$\Lambda_{(1)}^{0}(M)$.

The interior product of such a vector field $\mathbf{V}$ with a generalized
$p-$form $\mathbf{r}$ is a generalized $(p-1)-$form on $M$, denoted
$i_{\mathbf{V}}\mathbf{r}$ . \ Its definition is obtained by using the
approach of the previous section and extending the formulae there by
considering the action of the odd vector field%
\begin{equation}
i_{\mathbf{V}}=(v^{\rho}+v_{\sigma}^{\rho}\zeta^{\sigma}\mu\mathbf{)}%
\frac{\partial}{\partial\zeta^{\rho}}%
\end{equation}
on the function $\mathfrak{r}$ on\ $\widetilde{M}$ given in Eq.(31); that is
$i_{\mathbf{V}}\mathbf{r}$ on $M$ is defined to correspond to the function on
$\widetilde{M}$ given by%
\begin{equation}
i_{\mathbf{V}}\mathfrak{r}=(v^{\rho}+v_{\sigma}^{\rho}\zeta^{\sigma}%
\mu\mathbf{)}\frac{\partial}{\partial\zeta^{\rho}}[\frac{1}{p!}\rho
_{\alpha_{1}....\alpha_{p}}(x^{\alpha})\zeta^{\alpha_{1.....}}\zeta
^{\alpha_{p}}+\frac{1}{(p+1)!}\sigma_{\alpha_{1}....\alpha_{p+1}}(x^{\alpha
})\zeta^{\alpha_{1.....}}\zeta^{\alpha_{p+1}}\mu].
\end{equation}
It follows that on $M$ the interior product with respect to $\mathbf{V}$ is
given by the formula%
\begin{equation}
i_{\mathbf{V}}\mathbf{r}=\mathbf{v}^{\rho}i_{\frac{\partial}{\partial x^{\rho
}}}\mathbf{r.}%
\end{equation}
For $p$ equal to minus one and zero
\begin{align}
i_{\mathbf{V}}\overset{-1}{\mathbf{r}} &  =0.\\
i_{\mathbf{V}}\overset{0}{\mathbf{r}} &  =\sigma_{\alpha}v^{\alpha}%
\mathbf{m},\nonumber
\end{align}
and for $p\geqq1$%
\begin{align}
i_{\mathbf{V}}\mathbf{r} &  =i_{v}\mathbf{r+}\overset{p}{\gamma}%
\mathbf{m}=i_{v}\rho+i_{v}\sigma\mathbf{m+}\overset{p}{\gamma}\mathbf{m,}\\
\overset{p}{\gamma} &  =(-1)^{p-1}v_{\beta}^{\alpha}dx^{\beta}(i_{\frac
{\partial}{\partial x^{\alpha}}}\rho)\nonumber\\
&  =\frac{(-1)^{p-1}}{(p-1)!}v_{\lambda_{1}}^{\alpha}\rho_{\alpha\lambda
_{2}....\lambda_{p}}dx^{\lambda_{1}...}dx^{\lambda_{p}}.\nonumber
\end{align}
This interior product satisfies the graded Leibniz rule%
\begin{equation}
i_{\mathbf{V}}(\overset{p}{\mathbf{a}}\overset{q}{\mathbf{b}})=(i_{\mathbf{V}%
}\overset{p}{\mathbf{a}})\overset{q}{\mathbf{b}}+(-1)^{p}\overset
{p}{\mathbf{a}}(i_{\mathbf{V}}\overset{q}{\mathbf{b}}),
\end{equation}
but does not in general anti-commute because the interior product on
generalized zero forms need not be zero,%
\begin{equation}
(i_{\mathbf{W}}\mathbf{\circ}i_{\mathbf{V}}+i_{\mathbf{V}}\mathbf{\circ
}i_{\mathbf{W}})\mathbf{r}\mathbf{=}(-1)^{p-1}\{[v_{\beta}^{\alpha}w^{\beta
}+w_{\beta}^{\alpha}v^{\beta}](i_{\frac{\partial}{\partial x^{\alpha}}}%
\rho\}\mathbf{m,}%
\end{equation}
where $\mathbf{W}$=$(w^{\rho}+w_{\sigma}^{\rho}dx^{\sigma}\mathbf{m)}%
\frac{\partial}{\partial x^{\rho}}$. \ However if%
\begin{align}
\mathbf{V} &  \mathbf{=}v+i_{v}\Xi^{\alpha}\mathbf{m}\frac{\partial}{\partial
x^{\alpha}},\mathbf{W=}w+i_{w}\Xi^{\alpha}\mathbf{m}\frac{\partial}{\partial
x^{\alpha}}\\
\Xi^{\alpha} &  =\frac{1}{2}\Xi_{\beta\gamma}^{\alpha}dx^{\beta}dx^{\gamma
},\nonumber
\end{align}
where $v,w\in\mathcal{V}_{(0)}(M)$ and $\Xi$ is an ordinary vector-valued
two-form, then%
\begin{equation}
(i_{\mathbf{W}}\mathbf{\circ}i_{\mathbf{V}}+i_{\mathbf{V}}\mathbf{\circ
}i_{\mathbf{W}})=0.
\end{equation}

The Lie derivative of generalized forms with respect to a generalized
form-valued vector field $\mathbf{V,}$ which will be denoted
$\pounds _{\mathbf{V}},$ is defined by%
\begin{equation}
\pounds _{\mathbf{V}}=d\circ i_{\mathbf{V}}+i_{\mathbf{V}}\circ d.
\end{equation}

Calculation of the corresponding supercommutator on $\widetilde{M}$ gives the
even vector field $\pounds _{\mathbf{V}}\mathfrak{\ }$where%
\begin{align}
\pounds _{\mathbf{V}}\mathfrak{r} &  \mathfrak{=(}v^{\alpha}\frac{\partial
}{\partial x^{\alpha}}+\frac{\partial v^{\alpha}}{\partial x^{\beta}}%
\zeta^{\beta}\frac{\partial}{\partial\zeta^{\alpha}}-\epsilon v_{\beta
}^{\alpha}\zeta^{\beta}\frac{\partial}{\partial\zeta^{\alpha}})\mathfrak{r}\\
&  +(v_{\beta}^{\alpha}\zeta^{\beta}\mu\frac{\partial}{\partial x^{\alpha}%
}+\frac{\partial v_{\beta}^{\alpha}}{\partial x^{\gamma}}\zeta^{\gamma}%
\zeta^{\beta}\mu\frac{\partial}{\partial\zeta^{\alpha}})\mathfrak{r}.\nonumber
\end{align}

It follows that on $M$ the Lie derivative of a generalized $p-$form
$\mathbf{r}$ on with respect to a generalized form valued vector field
$\mathbf{V}=\mathbf{v}^{\alpha}\frac{\partial}{\partial x^{\alpha}}$ is%
\begin{align}
\pounds _{\mathbf{V}}\mathbf{r}  &  \mathbf{=(}d\mathbf{\circ i_{\mathbf{V}%
}+i_{\mathbf{V}}\circ}d)\mathbf{r}\\
&  =\mathbf{(v}^{\alpha}\frac{\partial}{\partial x^{\alpha}}+d(\mathbf{v}%
^{\alpha})i_{\frac{\partial}{\partial x^{\alpha}}})\mathbf{r},\nonumber
\end{align}
where $\mathbf{v}^{\alpha}\frac{\partial\mathbf{r}}{\partial x^{\alpha}}$
denotes the expression%
\begin{equation}
\mathbf{v}^{\alpha}\{\frac{1}{p!}\frac{\partial}{\partial x^{a}}[\rho
_{\alpha_{1}....\alpha_{p}}(x^{\alpha})]dx^{\alpha_{1.....}}dx^{\alpha_{p}%
}+\frac{1}{(p+1)!}\frac{\partial}{\partial x^{a}}[\sigma_{\alpha_{1}%
....\alpha_{p+1}}(x^{\alpha})]dx^{\alpha_{1.....}}dx^{\alpha_{p+1}}%
\mathbf{m}\}.
\end{equation}
Hence when $p=-1$ and $\mathbf{r}=\sigma\mathbf{m}$%
\begin{equation}
\pounds _{\mathbf{V}}\mathbf{r}\mathbf{=v}^{\alpha}\frac{\partial\sigma
}{\partial x^{\alpha}}\mathbf{m},
\end{equation}

when $p=0$ and $\mathbf{r}=\rho+\sigma\mathbf{m},$%
\begin{equation}
\pounds _{\mathbf{V}}\mathbf{r}\mathbf{=\pounds }_{v}\mathbf{\rho
+[\pounds }_{v}\mathbf{\sigma}+v_{\beta}^{\alpha}(\frac{\partial\rho}{\partial
x^{\alpha}}-\epsilon\sigma_{\alpha})dx^{\beta}]\mathbf{m},
\end{equation}
when $p\geqq1$ and $\mathbf{r}=\rho+\sigma\mathbf{m},$%
\begin{align}
\pounds _{\mathbf{V}}\mathbf{r}  &  \mathbf{=\pounds }_{v}\mathbf{\rho-}%
\frac{\epsilon}{(p-1)!}v_{\beta}^{\alpha}\rho_{\alpha\lambda_{2}...\lambda
_{p}}dx^{\beta}dx^{\lambda_{2}}...dx^{\lambda_{p}}\\
&  \mathbf{+[\pounds }_{v}\mathbf{\sigma}+\frac{(-1)^{p}}{p!}v_{\beta}%
^{\alpha}\frac{\partial}{\partial x^{\alpha}}\rho_{\lambda_{1}...\lambda_{p}%
}+\frac{(-1)^{p}}{(p-1)!}\frac{\partial v_{\beta}^{\alpha}}{\partial
x^{\lambda_{1}}}\rho_{\alpha\lambda_{2}...\lambda_{p}}\nonumber\\
&  -\frac{\epsilon}{p!}v_{\beta}^{\alpha}\sigma_{\alpha\lambda_{1}%
...\lambda_{p}}]dx^{\beta}dx^{\lambda_{1}}...dx^{\lambda_{p}}\mathbf{m}%
.\nonumber
\end{align}

It follows from Eqs.(41) and (45) that $\pounds _{\mathbf{V}}$ satisfies the
Leibniz rule and is a derivation of degree zero.

The Lie bracket of two generalized form-valued vector fields $\mathbf{V=v}%
^{\alpha}$ $\frac{\partial}{\partial x^{\alpha}}$and $\mathbf{W=w}^{\alpha}$
$\frac{\partial}{\partial x^{\alpha}}$ on $M$ is the generalized form-valued
vector field, $[\mathbf{V,W}]$, defined by the relation%
\begin{equation}
(\pounds _{\mathbf{V}}\pounds _{\mathbf{W}}-\pounds _{\mathbf{W}%
}\mathbf{\pounds _{\mathbf{V}})r=\pounds }_{[\mathbf{V,W]}}\mathbf{r.}%
\end{equation}
Then%
\begin{align}
\lbrack\mathbf{V,W}]  &  =[\mathbf{V,W}]^{\gamma}\frac{\partial}{\partial
x^{\gamma}}\\
\lbrack\mathbf{V,W}]^{\gamma}  &  =[v,w]^{\gamma}+\{v^{\beta}\frac{\partial
}{\partial x^{\beta}}w_{\alpha}^{\gamma}-w^{\beta}\frac{\partial}{\partial
x^{\beta}}v_{\alpha}^{\gamma}+w_{\beta}^{\gamma}\frac{\partial}{\partial
x^{\alpha}}v^{\beta}-v_{\beta}^{\gamma}\frac{\partial}{\partial x^{\alpha}%
}w^{\beta}\nonumber\\
&  +v_{\alpha}^{\beta}\frac{\partial}{\partial x^{\beta}}w^{\gamma}-w_{\alpha
}^{\beta}\frac{\partial}{\partial x^{\beta}}v^{\gamma}+\epsilon v_{\beta
}^{\gamma}w_{\alpha}^{\beta}-\epsilon w_{\beta}^{\gamma}v_{\alpha}^{\beta
}\}dx^{\alpha}\mathbf{m,}\nonumber
\end{align}
where $[v,w]^{\gamma}\frac{\partial}{\partial x^{\gamma}}$ is the ordinary Lie
bracket of $v$ and $w$. \ It follows from Eq.(52) that the Lie bracket
satisfies the Jacobi identity, if $\mathbf{U}$, $\mathbf{V}$ and $\mathbf{W}$
are generalized form-valued vector fields $\in\mathcal{V}_{(1)}(M)$%
\begin{equation}
\lbrack\mathbf{U,}[\mathbf{V,W}]]+[\mathbf{V,}[\mathbf{W,U}]]+[\mathbf{W,}%
[\mathbf{U,V}]]=0.
\end{equation}
All the operators $d$, $i_{\mathbf{V}}$ , \pounds $_{\mathbf{V}}$ and
$[\mathbf{V,W}]$ reduce to the usual operators when acting on ordinary forms
and vector fields.

Henceforth generalized form-valued vector fields will be referred to as type
$N$ vector fields with $N=1$ here. \ Ordinary vector fields are therefore type
$N=0$ vector fields.

In \cite{lah1} and \cite{chat} \ the concept of a generalized vector field was
introduced and explored. \ Such a vector field $V$ is determined by a pair
consisting \ of an ordinary vector field $v\in\mathcal{V}_{(o)}(M)$ and \ a
scalar field $v_{0\text{ }}$on $M$. it is straightfoward to see that such a
generalized vector field is a type $N=1$ vector field $\mathbf{V=}%
v+(v_{\sigma}^{\rho}dx^{\sigma}\mathbf{m)}\frac{\partial}{\partial x^{\rho}}$
when the special choice
\begin{equation}
v_{\beta}^{\alpha}=\delta_{\beta}^{\alpha}v_{0}.
\end{equation}
is made. \ Moreover if $V$ and $W$ are two generalized vector fields, the
inner product, $I_{V}$ , Lie derivative $\mathfrak{L}_{V}$ \ and Lie bracket
$\{V,W\}$ introduced in \cite{lah1} and \cite{chat} are the same as the inner
product $i_{\mathbf{V}}$, Lie Derivative $\pounds _{\mathbf{V}}$ and Lie
bracket $[\mathbf{V,W}]$ for the generalized form-valued vector fields
$\mathbf{V}=(v^{\rho}+v_{0}dx^{\rho}\mathbf{m)}\frac{\partial}{\partial
x^{\rho}}$ \ and $\mathbf{W}=(w^{\rho}+w_{0}dx^{\rho}\mathbf{m)}\frac
{\partial}{\partial x^{\rho}}$.

In concluding this section it should be noted that in \cite{lah1} and
\cite{chat} it was observed that it was not possible, in general, to define a
generalized vector field which was a Lie derivative, $\mathfrak{L}_{V}W$, by
using the equation%
\begin{equation}
\mathfrak{L}_{V}\circ i_{W}-i_{W}\circ\mathfrak{L}_{V}=i_{\mathfrak{L}_{V}W}.
\end{equation}
This result also applies to general vector fields $\in$ $\mathcal{V}_{(1)}(M)$
and is not surprising in the light of the failure, as was noted above, of
inner products to anti-commute. \ A modified Lie derivative,$\widehat
{\mathfrak{L}}_{V}$ of generalized forms which could be used in Eq.(56) to
define a (modified) Lie derivative of generalized vector fields,
$\widehat{\mathfrak{L}}_{V}W$, was introduced. That construction will not be
pursued here but the following general observation can be made.\ \ Eqs.(31)
and (32) suggests that for generalized forms the exterior derivative, $d$,
splits naturally into two exterior derivatives%
\begin{equation}
d=d_{(0)}+\epsilon d_{(1)}%
\end{equation}
where $d=d_{(0)}$ when $\epsilon=0$ and $d_{(1)}\mathbf{m}=1$, $d_{(1)}%
\alpha=0$ for any ordinary form $\alpha$. \ Similarly \ any type $N=1$ vector
field can be naturally written as the sum of two vector fields $v$ and
$\mathbf{V}_{(1)}$%
\begin{equation}
\mathbf{V}=v+(v_{\sigma}^{\rho}dx^{\sigma}\mathbf{m)}\frac{\partial}{\partial
x^{\rho}}=v+\mathbf{V}_{(1)}.
\end{equation}
Operators, such as the modified Lie derivative operator, can be constructed by
making use of these splittings. \ In fact the modified Lie derivative operator
of \cite{lah1} and \cite{chat} is given by%
\begin{equation}
\widehat{\mathfrak{L}}_{\mathbf{V}}\mathbf{r}=\pounds _{\mathbf{V}%
}\mathbf{r-(}d_{(0)}\circ i_{\mathbf{V}_{(1)}}+i_{\mathbf{V}_{(1)}}\circ
d_{(0)})\mathbf{r,}%
\end{equation}
where $\mathbf{V}=(v^{\rho}+v_{0}dx^{\rho}\mathbf{m)}\frac{\partial}{\partial
x^{\rho}}$.

These ideas are illustrated in the following simple examples. \ First an
example using "pure" type $N=1$ vector fields, for which $\mathbf{V}%
=(v_{\sigma}^{\rho}dx^{\sigma}\mathbf{m)}\frac{\partial}{\partial x^{\rho}%
}=\mathbf{V}_{(1)}.$ is given. \ Second Hamiltonian type $N=1$ vector fields
are briefly introduced and a simple special case is discussed.

\textbf{Example 1 }Pure type $N=1$ vector fields:

Let $J_{i}$ be three $\left(  1,1\right)  $ type tensor fields which satisfy,
as for example with hyperk\"{a}hler metrics, the conditions $J_{i}%
J_{j}=\varepsilon_{ijk}J_{k\text{ }}$ where $i,j,k$ range from one to three
and $\varepsilon_{ijk}$ is the totally skew symmetric Levi=Civita symbol. \ If
$\mathbf{V}_{i}=\mathbf{V}_{i(1)}=\frac{1}{2}J_{i\beta}^{\alpha}dx^{\alpha
}\mathbf{m}\frac{\partial}{\partial x^{\alpha}}$ then $[\mathbf{V}%
_{i},\mathbf{V}_{j}]=\epsilon\varepsilon_{ijk}\mathbf{V}_{k}$. Hence when
$\epsilon$ is zero the vector fields commute and when $\epsilon$ is non-zero
the Lie brackets of the three pure type $N=1$ vector fields $\mathbf{V}%
_{i}=\mathbf{V}_{i(1)}=\frac{1}{2\epsilon}J_{i\beta}^{\alpha}dx^{\alpha
}\mathbf{m}\frac{\partial}{\partial x^{\alpha}}$ satisfy the so(3) Lie algebra
condition, $[\mathbf{V}_{i},\mathbf{V}_{j}]=\varepsilon_{ijk}\mathbf{V}_{k}$.

\textbf{Example 2 }Type $N=1$ Hamiltonian vector fields

Let $M$ be an even dimensional manifold with $n=2l$ and with local coordinates
$\{x^{\alpha}\}$. \ Let $\mathbf{s}=\Omega+\Upsilon\mathbf{m}$ be a
generalized symplectic two -form on $M,$ that is a closed, non-degenerate
generalized two-form. \ Since $\mathbf{s}$ is required to be closed $\Omega$
and $\Upsilon$ must both be closed when $\epsilon=0$ and \ when $\epsilon$ is
non-zero $\Upsilon=\frac{1}{\epsilon}d\Omega$ and $\mathbf{s=}\Omega+\frac
{1}{\epsilon}d\Omega\mathbf{m}$. $\ $The two-form $\mathbf{s}$ is defined to
be non-degenerate if and only if $\Omega$ is non-degenerate. \ In this case
both the components of $\mathbf{s}$ and the ordinary two-form $\Omega$ are
invertible (written as square matrices) and $\Omega$ defines an isomorphism
between $TM$ and $T^{\ast}M$;\ if $\Omega=\frac{1}{2}\Omega_{\alpha\beta
}dx^{\alpha}dx^{\beta}$ then conventionally $\Omega^{\alpha\gamma}%
\Omega_{\beta\gamma}=\delta_{\beta}^{\alpha}$, \cite{libermann}. \ This
definition permits non-zero pure generalized form-valued vector field
solutions $\mathbf{W=W}_{(1)}$, here termed kernel vector fields, to the
equation%
\[
i_{\mathbf{W}}\mathbf{s=}0\mathbf{.}%
\]
Let $\mathbf{H=}h+k\mathbf{m}$ be a generalized zero-form, then $\mathbf{V}%
_{H}$ is by definition a generalized form-valued Hamiltonian vector field
corresponding to $\mathbf{H}$ when%
\[
i_{\mathbf{V}_{H}}\mathbf{s}=-d\mathbf{H.}%
\]
Such a Hamiltonian vector field is also a Hamiltonian vector field for the
generalized zero-form $\mathbf{H+}d\mathbf{L}$ where $\mathbf{L}$ is any
generalized minus one-form $l\mathbf{m}$. \ When $\ \mathbf{H\rightarrow
H+}d\mathbf{L}$, h$\rightarrow h+l$ and $k\rightarrow k+dl$.

Employing the decomposition of Eq.(58) and writing $\mathbf{V}_{H}%
=v_{H}+\mathbf{V}_{H(1)}$ the solutions of this equation, modulo arbitrary
kernel vector fields, are given in terms of components by%
\begin{align*}
v_{H}^{\alpha} &  =\Omega^{\alpha\beta}(\epsilon k_{\beta}-h,_{\beta}),\\
v_{H\beta}^{\alpha} &  =\Omega^{\gamma\alpha}(k_{[\beta,\gamma]}-\frac{1}%
{2}v_{H}^{\mu}\Upsilon_{\mu\beta\gamma}),
\end{align*}
where $\Upsilon=\frac{1}{3!}\Upsilon_{\alpha\beta}dx^{\alpha}dx^{\beta
}dx^{\gamma}$, partial differentiation is denoted by a comma and square
brackets denote the totally skew part. \ By Eq.(52), and the fact that
$\pounds _{\mathbf{V}_{H}}\mathbf{s=}0$, the Lie bracket of two generalized
form-valued Hamiltonian vector fields is also a generalized form-valued
Hamiltonian vector field.

A case where there is much simplification arises when the ordinary two-form
$\Omega$ is itself symplectic, $\mathbf{s=}\Omega$ , and only generalized
vector fields, defined in Eq.(55), are considered. \ Then there are no kernel
generalized vector fields. \ The consistency of the solutions above now
requires that $2v_{0}\Omega=dk$ and that, in dimension greater than two,
$v_{0}$ must be a constant. \ In symplectic coordinates $(p_{a},q^{a})$, with
$\Omega=dp_{a}dq^{a}$ and with the choices $v_{0}$ a constant and
$k=2v_{0}p_{a}dq^{a}$ , the Hamiltonian generalized vector field is%
\begin{align*}
\mathbf{V}_{H}  &  =v_{H}+\mathbf{V}_{H(1)}\\
&  =\frac{\partial h}{\partial p_{a}}\frac{\partial}{\partial q^{a}}%
-(\frac{\partial h}{\partial q^{a}}-2\epsilon v_{0}p_{a})\frac{\partial
}{\partial p_{a}}-\mathbf{m}v_{0}(dq^{a}\frac{\partial}{\partial q^{a}}%
+dp_{a}\frac{\partial}{\partial p_{a}}).
\end{align*}

Integral curves for $v_{H}$ satisfy the generalized Hamilton's equations%
\[
\frac{dq^{a}}{dt}=\frac{\partial h}{\partial p_{a}},\frac{dp_{a}}{dt}%
=-(\frac{\partial h}{\partial q^{a}}-2\epsilon v_{0}p_{a}).
\]

For example, if $h=\Sigma_{a=1}^{l}\frac{1}{2}[(q^{a})^{2}+(p_{a})^{2}]$ the
solutions of the generalized Hamilton's equations are determined by the
solutions of the differential equations%
\[
\frac{d^{2}}{dt^{2}}q^{a}-2\epsilon v_{0}\frac{d}{dt}q^{a}+q^{a}=0.
\]
The appearance of the damping (or anti-damping) term when $\epsilon$ is
non-zero appears to parallel the appearance of a mass term in field equations
when $\epsilon$ is non-zero, as for example in \cite{rob2}.

The ideas above can be straightforwardly extended to type $N\geqq2$
generalized forms and vector fields by considering functions on the Whitney
sum, $\widetilde{M}^{N}$ of $\Pi TM$ and a reverse parity $\mathbb{R}^{N}$
bundle over $M$, that is a trivial vector bundle with fibre $\mathbb{R}^{N}$
replaced by $\mathbb{R}^{0\mid N}$ . \ Natural local coordinates on
$\widetilde{M}^{N}$ are $(x^{\alpha},\zeta^{\alpha},\mu^{i})$, where $\mu^{i}$
$(i=1..N)$ are anti-commuting coordinates on $\mathbb{R}^{0\mid N}$. \ Type
$N$ generalized forms on $M$ correspond, in the obvious generalization of the
type $N=1$ case, to functions on $\widetilde{M}^{N}$ which are polynomial in
the anticommuting coordinates. \ The exterior product of type $N$ forms on $M$
corresponds to the product of such functions on $\widetilde{M}^{N}$. \ When
the exterior derivatives of the \ basis minus one-forms on $M$ are given by
$d\mathbf{m}^{i}=\epsilon^{i}$, where $\epsilon^{i}$ are constants, the
exterior derivative of type $N$ generalized forms on $M$ corresponds to the
action of the vector field $\zeta^{\alpha}\frac{\partial}{\partial x^{\alpha}%
}$ $+\epsilon^{i}$\ $\frac{\partial}{\partial\mu^{i}}$ on such functions on
$\widetilde{M}^{N}$. \ The interior product $i_{\mathbf{V}}$ of a generalized
type $N$ form by a type $N$ generalized vector field $\mathbf{V}%
=\mathbf{v}^{\alpha}\frac{\partial}{\partial x^{\alpha}}$, where the
components $\mathbf{v}^{\alpha}$ are type $N$ generalized zero-forms,
corresponds to the action of the vector field $\mathbf{v}^{\alpha}%
\frac{\partial}{\partial\zeta^{\alpha}}$ on the corresponding functions on
$\widetilde{M}^{N}$.

The ideas of the previous sections also extend straightforwardly to
generalized form-valued tensor fields and geometrical objects. Such an
extension is outlined in the next section.

\section{Type $N=1$ generalized affine and metric connections}

In this section the formalism above will be applied and the tensor calculus of
generalized affine connections and metrics, when $N=1$, will be outlined.
\ The definition of a generalized affine connection is the same as the
definition of an ordinary affine connection except that ordinary forms,
including zero forms, are replaced by generalized forms. \ If $\{U_{I}\}$ is a
covering of an $n-$dimensional manifold $M$ by coordinate charts, each with
coordinates $\{x_{I}^{\alpha}\}$ then a generalized affine connection
$\mathbf{A}$ is an assignment of a $n\times n$ matrix-valued generalized
one-form, with $\ (\mu,\nu)$ entry $\mathbf{A}_{I\nu}^{\mu}$, to each set
$U_{I}$ and such that on $U_{I}\cap U_{J}$ , for all $I$ and $J$,%
\begin{equation}
\mathbf{A}_{J\nu}^{\mu}=(G_{IJ}^{-1})_{\gamma}^{\mu}dG_{IJ}{}_{\nu}^{\gamma
}+(G_{IJ}^{-1})_{\gamma}^{\mu}\mathbf{A}_{I\lambda}^{\gamma}G_{IJ}{}_{\nu
}^{\lambda},
\end{equation}
where%
\begin{equation}
G_{IJ}{}_{\nu}^{\mu}=\frac{\partial x_{I}^{\mu}}{\partial x_{J}^{\nu}}.
\end{equation}

The curvature two-form $\mathbf{F}_{I}$ is the generalized form%
\begin{equation}
\mathbf{F_{I\nu}^{\mu}=dA}_{I\nu}^{\mu}+\mathbf{A}_{I\rho}^{\mu}%
\mathbf{A}_{I\nu}^{\rho},
\end{equation}
and under the transformation in Eq.(60)%
\[
\mathbf{F_{J\nu}^{\mu}=}(G_{IJ}^{-1})_{\gamma}^{\mu}\mathbf{F}_{I\lambda
}^{\gamma}G_{IJ}{}_{\nu}^{\lambda}.
\]
On any coordinate chart such as $U_{I}$ the connection one-form $\mathbf{A}%
_{I\nu}^{\mu}$ can be written as \
\begin{equation}
\mathbf{A}_{I\nu}^{\mu}=\alpha_{I\nu}^{\mu}+\beta_{I\nu}^{\mu}\mathbf{m,}%
\end{equation}
where $\alpha_{I\nu}^{\mu}$ and $\beta_{I\nu}^{\mu}$ are respectively ordinary
matrix valued one-forms and the curvature two-form is then
\begin{equation}
\mathbf{F}_{I\nu}^{\mu}=\mathcal{F}_{I\nu}^{\mu}+\epsilon\beta_{I\nu}^{\mu
}+D\beta_{I\nu}^{\mu}\mathbf{m,}%
\end{equation}
where%
\begin{align}
\mathcal{F}_{I\nu}^{\mu} &  =d\alpha_{I\nu}^{\mu}+\alpha_{I\rho}^{\mu}%
\alpha_{I\nu}^{\rho},\\
D\beta_{I\nu}^{\mu} &  =d\beta_{I\nu}^{\mu}+\alpha_{I\rho}^{\mu}\beta_{I\nu
}^{\rho}-\beta_{I\rho}^{\mu}\alpha_{I\nu}^{\rho}.\nonumber
\end{align}
It follows from the above that the locally defined ordinary one-forms
$\alpha_{I\nu}^{\mu}$ and curvature two-forms $\mathcal{F}_{I\nu}^{\mu},$
patch together to define global connection and curvature forms, $\alpha$ and
$\mathcal{F}$ , of an ordinary affine connection $D$. \ The ordinary two-forms
$\beta_{I\nu}^{\mu}$ transform as $\left(  1,1\right)  $ type tensor valued
two-forms. \ Henceforth connections on $M$ will be discussed and the
subscripts corresponding to coordinate charts will be dropped.

The curvature satisfies the Bianchi identities%
\begin{equation}
\mathbf{D\mathbf{F}_{\nu}^{\mu}=}d\mathbf{F}_{\nu}^{\mu}+\mathbf{A}_{\lambda
}^{\mu}\mathbf{F}_{\nu}^{\lambda}-\mathbf{F}_{\lambda}^{\mu}\mathbf{A}_{\nu
}^{\lambda}=0,
\end{equation}
\ where here $\mathbf{D}$ denotes the covariant exterior derivative of a type
$N=1$ valued generalized form. \ For a $\binom{1}{1}-$tensor valued
generalized p-form $\mathbf{P}$%
\begin{equation}
\mathbf{DP_{\nu}^{\mu}=}d\mathbf{P}_{\nu}^{\mu}+\mathbf{A}_{\lambda}^{\mu
}\mathbf{P}_{\nu}^{\lambda}+(-1)^{p+1}\mathbf{P}_{\lambda}^{\mu}%
\mathbf{A}_{\nu}^{\lambda}.
\end{equation}
The covariant derivative of a generalized zero-form is the exterior
derivative. \ If $\mathbf{V=v}^{\rho}\frac{\partial}{\partial x^{\rho}%
}=(v^{\rho}+v_{\sigma}^{\rho}dx^{\sigma}\mathbf{m)}\frac{\partial}{\partial
x^{\rho}}$, the covariant derivative is%
\begin{equation}
\mathbf{\nabla V}=\mathbf{Dv}^{\mu}\mathbf{\otimes}\frac{\partial}{\partial
x^{\mu}}=(d\mathbf{v}^{\mu}+\mathbf{A}_{\nu}^{\mu}\mathbf{v}^{\nu
}\mathbf{)\otimes}\frac{\partial}{\partial x^{\mu}},
\end{equation}
where%
\begin{equation}
\mathbf{Dv}^{\mu}\mathbf{=}Dv^{\mu}-\epsilon v_{\nu}^{\mu}dx^{\nu}+[D(v_{\nu
}^{\mu}dx^{\nu})+\beta_{\nu}^{\mu}v^{\nu}]\mathbf{m,}%
\end{equation}
and $\mathbf{D}$ and $D$ are the covariant exterior derivatives with respect
to $\mathbf{A}$ and $\alpha$ respectively. \ The covariant derivative with
respect to a type $N=1$ vector field $\mathbf{W}$ is the generalized
form-valued vector field%
\begin{equation}
\mathbf{\nabla}_{\mathbf{W}}\mathbf{V=[i}_{\mathbf{W}}((d\mathbf{v}^{\alpha
}+\mathbf{A}_{\beta}^{\alpha}\mathbf{v}^{\beta}\mathbf{)]}\frac{\partial
}{\partial x^{\alpha}}.
\end{equation}
The covariant derivative is extended to type $N=1$ generalized form -valued
tensor fields by using the linearity and product rules satisfied by ordinary
covariant derivatives and tensor fields.

A field $\mathbf{V}$ is a parallel vector field if $\mathbf{\nabla V}=0$, that is%

\begin{align}
Dv^{\mu}-\epsilon v_{\nu}^{\mu}dx^{\nu}  &  =0,\\
D(v_{\nu}^{\mu}dx^{\nu})+\beta_{\nu}^{\mu}v^{\nu}  &  =0.\nonumber
\end{align}
and such a system of equations is completely integrable if and only if the
generalized curvature $\mathbf{\mathbf{F}}$ is zero, that is when%
\begin{align}
\mathcal{F}_{\nu}^{\mu}  &  =-\epsilon\beta_{\nu}^{\mu}.\\
D\beta_{\nu}^{\mu}  &  =0.\nonumber
\end{align}

A generalized metric can be defined by straightforwardly extending the
definition of an ordinary metric to encompass generalized forms. \ A
generalized metric $\mathbf{g}$ is a smooth symmetric bilinear function on
$\mathcal{V}_{(1)}$ $_{p}(M)$ at each $p\in M$%
\begin{align}
\mathbf{g(V,W)}  &  \mathbf{=g}_{\mu\nu}\mathbf{v}^{\mu}\mathbf{w}^{\nu},\\
\mathbf{g}_{\mu\nu}  &  =\mathbf{g}_{v\mu},\nonumber
\end{align}
where the components are generalized zero-forms and $\mathbf{V}$ and
$\mathbf{W}$ are any type $N=1$ vector fields as above. \ If%
\begin{equation}
\mathbf{g}_{\mu\nu}=\gamma_{\mu\nu}+\chi_{\mu\nu}\mathbf{m,}%
\end{equation}
where $\gamma_{\mu\nu}$ and $\chi_{\mu\nu}$ are ordinary zero and one-forms
respectively, the generalized metric is said to be non-degenerate when
$det(\gamma_{\mu\nu})$ is non-zero. \ A non-degenerate metric has inverse%
\begin{equation}
\mathbf{g}^{_{\mu\nu}}=\gamma^{_{\mu\nu}}-\chi^{_{\mu\nu}}\mathbf{m}%
\end{equation}
where $\gamma^{_{\mu\nu}}\gamma_{\nu\rho}=\delta_{\rho}^{\mu}$ and
$\chi^{_{\mu\nu}}=\gamma^{_{\mu\rho}}\gamma^{_{\mu\sigma}}\chi_{\rho\sigma}$.
\ Henceforth only non-degenerate metrics will be considered.

The expanded form of $\mathbf{g}_{\mu\nu}\mathbf{v}^{\mu}\mathbf{w}^{\nu}$ is
\begin{equation}
\mathbf{g}_{\mu\nu}\mathbf{v}^{\mu}\mathbf{w}^{\nu}=\gamma_{\mu\nu}v^{\mu
}w^{\nu}+(v_{\mu}w_{\rho}^{\mu}+w_{\mu}v_{\rho}^{\mu}+\chi_{\mu\nu\rho}v^{\mu
}w^{\nu})dx^{\rho}\mathbf{m,}%
\end{equation}
where here and henceforth indices are lowered (and raised) by using
$\gamma_{\mu\nu}$ and its inverse and $\chi_{\mu\nu}=\chi_{\mu\nu\rho}%
dx^{\rho}$.

If $\mathbf{A}$ is a generalized connection and $\mathbf{g}$ then $\mathbf{A}$
is a generalized metric connection when the covariant derivative of
$\mathbf{g}$ is zero. \ This compatibility condition may be expressed as the
vanishing of the generalized non-metricity one-form $\mathbf{Q}_{_{\mu\nu}}$
where%
\begin{equation}
\mathbf{Dg}_{\mu\nu}=d\mathbf{g}_{\mu\nu}-\mathbf{g}_{\mu\lambda}%
\mathbf{A}_{\nu}^{\lambda}-\mathbf{g}_{\lambda\nu}\mathbf{A}_{\mu}^{\lambda
}=\mathbf{Q}_{\mu\nu}.
\end{equation}
Here
\begin{align}
\mathbf{Q}_{\mu\nu}  &  =q_{_{\mu\nu}}-\epsilon\chi_{\mu\nu}+[D\chi_{\mu\nu
}-(\beta_{\mu\nu}+\beta_{\nu\mu})]\mathbf{m},\\
D\chi_{\mu\nu}  &  =d\chi_{\mu\nu}-\alpha_{\mu}^{\lambda}\chi_{\lambda\nu
}-\alpha_{\nu}^{\lambda}\chi_{\mu\lambda,}\nonumber
\end{align}
and $q_{\mu\nu}$ is the non-metricity one-form for $\alpha_{\nu}^{\mu}$ and
$\gamma_{\mu\nu}$,%
\begin{equation}
q_{_{\mu\nu}}=D\gamma_{\mu\nu}=d\gamma_{\mu\nu}-\gamma_{\mu\lambda}\alpha
_{\nu}^{\lambda}-\gamma_{\lambda\nu}\alpha_{\mu}^{\lambda}.
\end{equation}
When $\epsilon=0$, $\mathbf{Q}_{_{\mu\nu}}=0$ if and only if%
\begin{align}
q_{_{\mu\nu}}  &  =0,\\
D\chi_{\mu\nu}  &  =(\beta_{_{\mu\nu}}+\beta_{\nu\mu}),\nonumber
\end{align}
that is $\alpha$ is a metric connection for the metric $\gamma_{\mu\nu}%
dx^{\mu}dx^{\nu}$ and%
\begin{equation}
\mathbf{A}_{\nu}^{\mu}=\alpha_{\nu}^{\mu}+(\widetilde{\beta}_{.\nu}^{\mu
}+\frac{1}{2}D\chi_{\nu}^{\mu})\mathbf{m}%
\end{equation}
where $\widetilde{\beta}_{\nu}^{\mu}=\frac{1}{2}\gamma^{\mu\lambda}%
(\beta_{\lambda\nu}-\beta_{\nu\lambda})$.

When $\epsilon\neq0$, $\mathbf{Q}_{_{\mu\nu}}=0$ if and only if%
\begin{align}
\mathbf{g}_{\mu\nu}  &  =\gamma_{\mu\nu}+\epsilon^{-1}q_{_{\mu\nu}}%
\mathbf{m,}\\
\mathbf{A}_{\nu}^{\mu}  &  =\alpha_{\nu}^{\mu}+[\widetilde{\beta}_{\nu}^{\mu
}-\frac{1}{2\epsilon}(\mathcal{F}_{.\nu}^{\mu}+\mathcal{F}_{\nu}^{.\mu
})]\mathbf{m.}\nonumber
\end{align}
Hence there is the following extension of the fundamental theorem of
Riemannian geometry.

Let $\mathbf{g}_{\mu\nu}=\gamma_{\mu\nu}+\chi_{\mu\nu}\mathbf{m}$ be a
generalized metric. \ Then if $\mathbf{A}_{\nu}^{\mu}=\alpha_{\nu}^{\mu}%
+\beta_{\nu}^{\mu}\mathbf{m}$ is a generalized connection where $\alpha_{\nu
}^{\mu}$ has zero torsion and $\beta_{_{\mu\nu}}=\beta_{\nu\mu}$:

(i) When $\epsilon=0$ the only such connection which is metric, that is
$\mathbf{Dg}_{\mu\nu}=0$, is $\mathbf{A}_{\nu}^{\mu}=\alpha_{\nu}^{\mu}%
+\frac{1}{2}D\chi_{\nu}^{\mu}\mathbf{m}$ where $\alpha_{\nu}^{\mu}$ is the
unique Levi-Civita connection for the metric $\gamma_{\mu\nu}dx^{\mu}dx^{\nu}%
$. \ In this case the generalized curvature form is $\mathbf{F}_{\nu}^{\mu
}=\mathcal{F}_{\nu}^{\mu}+\frac{1}{2}(\mathcal{F}_{\lambda}^{\mu}\chi_{\nu
}^{\lambda}-\chi_{\lambda}^{\mu}\mathcal{F}_{\nu}^{\lambda})\mathbf{m}$.

(ii) When $\epsilon\neq0$ the only such connections which are metric are given
by $\mathbf{A}_{\nu}^{\mu}=\alpha_{\nu}^{\mu}-\frac{1}{2\epsilon}%
(\mathcal{F}_{.\nu}^{\mu}+\mathcal{F}_{\nu}^{.\mu})\mathbf{m}$ with curvature
two-forms $\mathbf{F}_{\nu}^{\mu}=\frac{1}{2}(\mathcal{F}_{.\nu}^{\mu
}-\mathcal{F}_{\nu}^{.\mu})-\frac{1}{2\epsilon}(q_{\nu\lambda}\mathcal{F}%
^{\lambda\mu}+q^{\mu\lambda}\mathcal{F}_{\nu\lambda})\mathbf{m}$.

Note that in the latter case if the generalized metric is an ordinary metric,
that is $\mathbf{g}_{\mu\nu}=\gamma_{\mu\nu}$, then the only such generalized
connection which is metric is $\mathbf{A}_{\nu}^{\mu}=\alpha_{\nu}^{\mu}$ with
generalized curvature $\mathbf{F}_{\nu}^{\mu}=\mathcal{F}_{.\nu}^{\mu}$, where
$\alpha_{\nu}^{\mu}$ is the unique Levi-Civita connection for the metric
$\gamma_{\mu\nu}dx^{\mu}dx^{\nu}$.

\section{Discussion}

In this paper type $N=1$ generalized form-valued vector fields have been
constructed and it has been shown that generalized vector fields constitute a
sub-class of such fields \ \ Generalized affine connections and metrics have
also been introduced. \ It is a straightforward matter to extend the results
in this paper to general vector bundles, generalized form-valued sections and
generalized connections. \ The latter, discussed in earlier papers, bear a
formal similarity to connections used in the higher gauge theories reviewed in
\cite{baez}. Those generalized connections have been used to formulate
Lagrangian field theories and a similar use can be made of the generalized
affine connections and metrics introduced here.

\textbf{Acknowledgement: }I would like to thank Alice Rogers for some useful discussions.

\section{Appendix: Exterior derivatives of type N=1 forms}

The exterior derivative $d:\Lambda_{(0)}^{p}(M)\rightarrow\Lambda_{(0)}%
^{p+1}(M)$ for ordinary forms is uniquely determined by the four conditions
\cite{chern}

(i) $d(\alpha+\beta)=d\alpha+d\beta$.

(ii) for $f\in f\in\Lambda_{(0)}^{0}(M)$, $df$ has its usual meaning as the
differential of $f$,

(iii) $d\circ d=0,$

(iv) $d(\alpha\beta)=d\alpha\beta+(-1)^{p}\alpha d\beta$, where $\alpha$ is a
$p-$form.

Exterior derivatives, $d:\Lambda_{(N)}^{p}(M)\rightarrow\Lambda_{(N)}%
^{p+1}(M)$, for generalized forms of type $N$ greater than zero, also satisfy
these conditions but they are not uniquely determined by them. \ The aim of
this appendix is to discuss this point by developing previous work,
\cite{rob5}, and constructing global solutions of the differential ideal
defining the exterior derivative. \ This will be done here for type $N=1$
forms since they can be treated most easily and completely.

Let $d:\Lambda_{(1)}^{p}(M)\rightarrow\Lambda_{(1)}^{p+1}(M)$ be an exterior
derivative for type $N=1$ forms on an on a real $n-$dimensional differentiable
manifold $M$. \ Assume that $\mathbf{m}$ is a non-zero minus one-form and that
any type $N=1$ generalized $p-$form, $\overset{p}{\mathbf{a}}\in\Lambda
_{(1)}^{p}(M)$, may be expressed as%
\begin{equation}
\overset{p}{\mathbf{a}}=\overset{p}{\alpha}+\overset{p+1}{\alpha}\mathbf{m},
\end{equation}
\ where the ordinary forms $\overset{p}{\alpha}$ and $\overset{p+1}{\alpha}$
are respectively of degree $p$ and $p+1$ on $M$, and $p$ can take integer
values from $-1$ to $n$. \ All the forms are assumed to obey the usual rules
of exterior algebra and calculus and the exterior derivative of $\mathbf{m}$
is required to be a type $N=1$ generalized zero-form.

It follows that%
\begin{equation}
d\mathbf{m}=\vartheta-\varphi\mathbf{m}%
\end{equation}
where $\vartheta$ is an ordinary zero-form and $\varphi$ is an ordinary
one-form on $M$. \ Then $d^{2}\mathbf{m}=0$ if and only if%
\begin{align}
d\vartheta+\vartheta\varphi &  =0,\\
d\varphi &  =0.\nonumber
\end{align}
The solutions of this closed differential ideal of ordinary forms determine
the possible exterior derivatives $d$. \ The exterior derivative of any type
$N=1$ form $\overset{p}{\mathbf{a}}$ is then given by%
\begin{equation}
d\overset{p}{\mathbf{a}}=d\overset{p}{\alpha}+(-1)^{p+1}\vartheta\overset
{p+1}{\alpha}+[d\overset{p+1}{\alpha}-\varphi\overset{p+1}{\alpha}]\mathbf{m.}%
\end{equation}

Consider now the consequences of these global assumptions. \ In a contractible
open set $U$ on $M$ the closed form $\varphi$ is exact. \ Therefore in $U$,%
\begin{align}
\varphi &  =d\xi,\\
\vartheta &  =\tau\exp(-\xi),\nonumber
\end{align}
for some constant $\tau$ and some function $\xi$. \ Hence, in $U$%
\begin{equation}
d\mathbf{m}=\tau\exp(-\xi)-d\xi\mathbf{m.}%
\end{equation}
The pair $(\tau$,$\xi)$ is not unique since there is the freedom
$\tau\rightarrow\tau\exp\chi$, $\xi\rightarrow\xi+\chi$, where $\chi$ is a constant.

Consider next a good covering of $M$ by a family of (contractible) open sets
$\{U_{I}\}$. \ By Eqs.(87) and (88) there are constants and functions
$(\tau_{I}$, $\xi_{I})$ such that%
\begin{align}
\varphi &  =d\xi_{I},\\
\vartheta &  =\tau_{I}\exp(-\xi_{I}),\nonumber\\
d\mathbf{m}  &  =\tau_{I}\exp(-\xi_{I})-d\xi_{I}\mathbf{m,}\nonumber
\end{align}
on $U_{I}$ and similarly on each set in the covering. \ On the intersection of
any two sets in the covering, $U_{I}$ and $U_{J}$ say, it follows from Eq.(89)
that
\begin{align}
\tau_{I}\exp(-\xi_{I})  &  =\tau_{J}\exp(-\xi_{J}),\\
d\xi_{I}  &  =d\xi_{J}.\nonumber
\end{align}
Hence on any intersection such as $U_{I}\cap U_{J}$%

\begin{align}
\xi_{I}-\xi_{J}  &  =\tau_{IJ},\\
\tau_{I}  &  =\tau_{J}\exp\tau_{IJ},\nonumber
\end{align}
for constants $\tau_{IJ}$ satisfying $\tau_{IJ}=-\tau_{JI}$. \ Consistency on
triple intersections, $U_{I}\cap U_{J}\cap U_{K}$ requires that%
\begin{equation}
\tau_{IJ}+\tau_{JK}+\tau_{KI}=0.
\end{equation}
Therefore, on $U_{I}$%
\begin{equation}
d\overset{p}{\mathbf{a}}=d\overset{p}{\alpha}+(-1)^{p+1}\tau_{I}\exp(-\xi
_{I})\overset{p+1}{\alpha}+[d\overset{p+1}{\alpha}-d\xi_{I}\overset
{p+1}{\alpha}]\mathbf{m,}%
\end{equation}
and similarly on all the sets in the open covering.

From Eq.(90) it follows that if $\tau_{I}$ is zero so is $\tau_{J}$ and then
on each set $U_{I}$ in the open cover%
\begin{equation}
d\mathbf{m}=-d\xi_{I}\mathbf{m}%
\end{equation}
and $\vartheta=0$ on $M$. \ Call this case (i). \ On the other hand if
$\tau_{I}$ is non-zero in $U_{I}$ then, from Eq.(90) $\tau_{J}$ is non-zero in
$U_{J}$ and hence $\vartheta$ must be non-zero in $M$. \ Call this case (ii).

Now consider rescalings of $\mathbf{m}$ and $\overset{p+1}{\alpha}$. \ On each
open set of the cover such as $U_{I}$ let $c_{I}$ be a non-zero constant, and
on any intersection such as $U_{I}\cap U_{J}$ let these constants be related
by%
\begin{equation}
c_{I}=c_{J}\exp\tau_{IJ}.
\end{equation}
On $U_{I}$ let%
\begin{align}
\widetilde{\mathbf{m}}_{I}  &  =c_{I}^{-1}\exp(\xi_{I})\mathbf{m,}\\
\overset{p+1}{\widetilde{\alpha}}_{I}  &  =c_{I}\exp(-\xi_{I})\overset
{p+1}{\alpha},
\end{align}
with similar scalings on the other sets in the open cover. \ Then%
\begin{align}
\overset{p}{\mathbf{a}}  &  =\overset{p}{\alpha}+\overset{p+1}{\widetilde
{\alpha}}_{I}\widetilde{\mathbf{m}}_{I},\\
d\widetilde{\mathbf{m}}_{I}  &  =\tau_{I}c_{I}^{-1},\nonumber\\
d\overset{p}{\mathbf{a}}  &  =[d\overset{p}{\alpha}+(-1)^{p+1}\tau_{I}%
c_{I}^{-1}\overset{p+1}{\widetilde{\alpha}}_{I}]+d\overset{p+1}{\widetilde
{\alpha}}_{I}\widetilde{\mathbf{m}}_{I},\nonumber
\end{align}
on $U_{I}$ and similarly on all the sets in the cover. \ It follows from
Eq.(90) and the following equations that on any intersection such as
$U_{I}\cap U_{J}$%
\begin{align}
\widetilde{\mathbf{m}}_{I}  &  =\widetilde{\mathbf{m}}_{J},\text{ }%
\overset{p+1}{\widetilde{\alpha}}_{I}=\overset{p+1}{\widetilde{\alpha}}_{J},\\
d\widetilde{\mathbf{m}}_{I}  &  =d\widetilde{\mathbf{m}}_{J},\text{ }%
d\overset{p+1}{\widetilde{\alpha}}_{I}=d\overset{p+1}{\widetilde{\alpha}}%
_{J},\nonumber
\end{align}
and there is consistency on triple intersections.

In case (i), for all $c_{I}$%
\begin{align}
d\widetilde{\mathbf{m}}_{I}  &  =0,\\
d\overset{p}{\mathbf{a}}  &  =d\overset{p}{\alpha}+d\overset{p+1}%
{\widetilde{\alpha}}_{I}\widetilde{\mathbf{m}}_{I}%
\end{align}

In case (ii), making the choice of constants $c_{I}$%
\begin{equation}
c_{I}=\tau_{I}\epsilon^{-1}%
\end{equation}
in $U_{I}$ , where $\epsilon$ is a real non-zero constant, gives%
\begin{align}
d\widetilde{\mathbf{m}}_{I}  &  =\epsilon,\\
d\overset{p}{\mathbf{a}}  &  =[d\overset{p}{\alpha}+(-1)^{p+1}\epsilon
\overset{p+1}{\widetilde{\alpha}}_{I}]+d\overset{p+1}{\widetilde{\alpha}}%
_{I}\widetilde{\mathbf{m}}_{I},\nonumber
\end{align}
and similarly for all the open sets in the cover. \ When $\epsilon=1$ this
choice corresponds to the choice of what has been termed a canonical basis on
$M.$

\end{document}